\documentclass[letter,scriptaddress,twocolumn]{revtex4}
\usepackage{float} 
\usepackage{amsmath}

\def\beq{\begin{equation}\begin{aligned}}
\def\eeq{\end{aligned}\end{equation}}
  
\usepackage{amscd,amssymb,latexsym,subfigure,verbatim,hyperref,epsfig,color}
 \usepackage{indentfirst}

\begin{document}
\title{  COVID-19 Forecasts via Stock Market Indicators}
\author{Yi Liang$^{\star}$
and James Unwin$^{\dagger}$ }
  \affiliation{$\star$~PRIMES, Massachusetts Institute of Technology, Cambridge, MA 02139, USA\\
$\dagger$~Department of Physics,  University of Illinois at Chicago, Chicago, IL 60607, USA}

\renewcommand{\thesection}{\Roman{section}}
\begin{abstract}
Reliable short term forecasting can provide potentially lifesaving insights into logistical planning, and in particular, into the optimal allocation of resources such as hospital staff and equipment. By reinterpreting COVID-19 daily cases in terms of candlesticks, we are able to apply some of the most popular stock market technical indicators  to obtain predictive power over the course of the  pandemics. By providing a quantitative assessment of MACD, RSI, and candlestick analyses, we show their statistical significance  in making predictions for both stock market data and WHO COVID-19 data. In particular, we show the utility of this novel approach by considering the identification of the beginnings of subsequent waves of  the pandemic. Finally, our new methods are used to assess whether current health policies are impacting the growth in new COVID-19 cases.
  \end{abstract}
\maketitle

\section{Introduction}

Logistical planning can be the difference between life and death
during a pandemic, such as the ongoing COVID-19 crisis. Here we
identify new techniques which can be applied during pandemics to
assist  in the optimal allocation of resources, and to aid in the
evaluation of current health policies. Specifically, we repurpose a
number of tools developed as stock market strategies and demonstrate
that these techniques can be applied to predict future trends in the
number of daily new cases of COVID. Notably, these tools can be used
to \begin{itemize}
\item Identify the peak of a pandemic wave;
\item Forecast the start of new waves.
\end{itemize}

While fluctuations in the number of new COVID cases and the prices of
stocks may naively seem disconnected, both systems can be described as
\textit{\textbf{non-stationary random walks}}, i.e.~a time series which exhibits random
fluctuations around a longer-term trend. In the context of stocks, the random walk
hypothesis can be formulated with the daily rate of returns in the
stock market randomly drawn from a Gaussian or Laplace distribution
\cite{randomwalk}. On the other hand, the daily increase in COVID-19 cases can be modeled as a random 
walk due to the complex nature of human interactions and it has an overall trend as 
an infectious disease in the spread and controlled phases. Drawing on this connection 
 we hypothesized that strategies developed for predicting stock price movements can be
repurposed to forecast changes in the trend of the number of new COVID
cases, and more generally any system that is well described as a non-stationary random walk. 
In this work, we show that these techniques, collectively known
as ``technical analysis'', do indeed provide accurate predictions of
the COVID-19 pandemic, in the sense that they are statistically
significant.

Traders look to identify
continuations or reversals in stock market trends to profit from
short-term trades, and there are several well established set of
techniques for forecasting future stock movements, known as
``technical indicators''.  Popular technical indicators include
candlestick patterns  \cite{bulkowski1}, the Moving Average
Convergence Divergence (MACD) indicator  \cite{macd1}, and the
Relative Strength Index (RSI) indicator \cite{rsi1}.  While, the
robustness of these techniques has been debated \cite{prof} -- in
particular,  the efficient market hypothesis states that if the market
is efficient, then any profitability information of technical
indicators would be incorporated into the new prices and thus it
should be impossible to gain abnormal profits \cite{emh} -- it has
been shown that several of these technical indicators are indeed
statistically significant.  Specifically, statistical tests of
technical indicators have been conducted to examine the effectiveness,
and the profitability of  the stock markets of Taiwan, Thailand,
USA, and Brazil \cite{profit_taiwan,profit_thai,profit_brazil,
profit_1, macd2}. These studies have 
shown that some techniques do have predictive power, providing
evidence for market inefficiency, while others indicators are not
successful.

After introducing the notion of {\it trends} and {\it candlestick representations} for time series data, we  provide mathematical definitions for the various technical indicators  in Section \ref{Sec2}.  To emphasize the novelty of our techniques, we use the World Health Organization (WHO) COVID-19 data throughout the paper and highlight the   application of these
  technical analysis techniques outside of the domain of the
financial markets. By performing statistical tests on technical indicators in Section \ref{Sec3}, we show that a
selection of these indicators correctly predict reversals in existing
trends at rates which are statistically significant. Our analysis
improves on aspects of earlier analyses of stock market data and,
moreover, extends the study to these indicators outside of stock
market forecasts, as illustrated in Section \ref{Sec4} for COVID-19. Our results have the following important
implications:
\begin{itemize}
\item Technical indicators have predictive power beyond forecasting
future asset prices.
\item These tools can be used to identify the beginnings of subsequent
waves of  a pandemic.
\item New methods to asses whether current health policies are
impacting the growth in new cases.
\end{itemize}
For completeness, Section \ref{Sec5} gives an analysis using stock market data. Section~\ref{Sec6} gives some concluding remarks. 

\section{Technical Indicators}
\label{Sec2}

Technical analysis aims to forecast future price movements in the stock market, and will be the key tools which we shall use to analyze COVID-19 data. In this section we first introduce the notion of {\it a trend}  (Section \ref{sec1.1}) and {\it Candlestick Representations} of data (Section \ref{candRep}), and then outline some of the main technical indicators:
\begin{itemize}
\item Candlestick Patterns (Section \ref{cand_pat})
\item RSI  (Section \ref{RSI_sec})
\item MACD  (Section \ref{macd_sec})
\end{itemize}
 In subsequent sections, we will investigate the predictive power of these technical indicators on COVID-19 data.

\subsection{Trends}
\label{sec1.1}

\begin{center}
 \begin{figure}[t]
	\includegraphics[width= 0.45\textwidth]{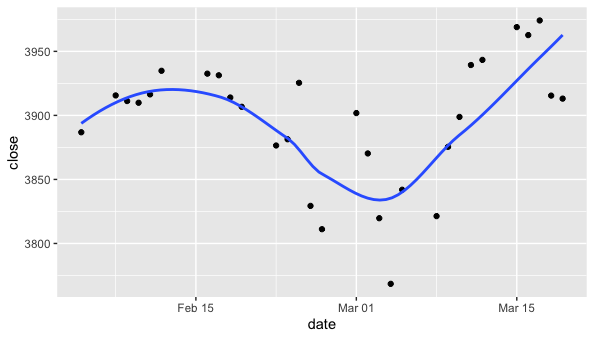}
	\caption{Daily Prices for S\&P 500 from February 5, 2020 to March 19, 2020 with our simple polynomial regression line (shown in blue) indicates the trend. Observe that the price follows the general trend for a few unit of time, and thus a regression line  can provide a useful insight. \label{trend}}	
\end{figure}
\end{center}

 While stock prices and daily COVID cases may fluctuate on shorter time frames, they have an observed tendency to evolve in the same direction for extended periods of time. The cause for long term trends in the stock market may be linked to macroeconomic factors such as monetary policy, or in the case of individual companies trends may be due to particular news or sentiment which result in the continual increases or decreases of the stock prices.   For  COVID-19, the growth of new cases is due to the fact that the virus is very infectious and the population was (and remains) highly vulnerable, which led to a significant initial uptrend.  
As a result, there is typically a well defined notion of ``trends'' in time series such as these. A good approach to identify trends in a time series is  through a simple regression procedure (as we will detail below) and Figure \ref{trend} presents an example in terms of the daily change in asset prices.

A period for which the gradient of the regression line has a fixed sign indicates a \textbf{trend} in the time series, which can be either an uptrend or a downtrend. 
Identifying such trends can provide insight into the likely subsequent behavior of the time series. Technical analysis was originally developed to provide signals for the start and end of trends in stock market data, and here we apply these techniques to alternative time series data sets.

We introduce a time variable which is integer valued and in particular we focus on daily and weekly intervals, counting from some partiular start date. A natural question to ask is whether for a given date $D$ a particular evolution in the time series exhibits a preexisting trend over the last $\delta$ days. If such a trend exists we will say that the data exhibits a $\delta$-interval trend, additionally: 
\begin{itemize}
\item  A trend is called \textit{\textbf{bullish}} if the gradient of the trend is strictly positive.
\item  Conversely, a trend is said to be \textit{\textbf{bearish}} its  gradient is strictly negative.
\end{itemize}

In what follows we shall often partition the time series data into  daily and weekly intervals.  Over the course of an interval the value of the time series will vary, following conventions of stock market analyses, we shall track a number of characteristic features of each time interval, in particular: 
\begin{itemize}
\item The initial value for a given time interval,  is called the {\bf opening value} and is denoted $O_t$.
\item  The final value for a given time interval,  is called the {\bf closing value} and is denoted $C_t$.
\end{itemize}
 The subscript $t$ is the index of the time interval.
 Since the data is discretized, typically $C_t\neq O_{t+1}$. It is also useful to define an average value for a given time interval
\begin{equation}
	M_t := \frac{O_t+C_t}{2}~.
\end{equation}

To identify potential  trends we apply a linear regression fitted to $M_t$ over the range of dates  $D-\delta\leq t \leq D$, with residuals $\gamma_t$ of the regression defined by
\begin{equation}
\gamma_t := {\rm abs}[(l(t)-M_t] ,
\end{equation}
where $l(t)$ is the value of the linear regression at time $t$.
We then define a trend function $T(\cdot)$ which takes the set of $\{M_t\}$ as input and returns $+1$ for an uptrend and $-1$ for an downwards  trend, as follows 
\begin{equation}
T(\cdot):= 
\begin{cases}
~1, & k \geq 0.005\cdot \mu , \qquad \gamma_t<0.02 \cdot \mu \\
-1,  &k \leq -0.005\cdot \mu, \quad \gamma_t<0.02 \cdot \mu \\
~0, & {\rm otherwise}
\end{cases}~~,
\end{equation}
 where $k$ denotes the slope of the regression and the mean is given by $\mu =  {\rm mean}(\{M_t\})$. 
The requirement  on $k$ corresponds to an increase or decrease of at least half of a percent of the mean. The restriction on $\gamma$ evaluates the goodness of the linear regression fit, requiring that each $\{M_t\}$ be no further than $2\%$ from the trend line.  The function returns zero if there is not a robust trend in values, indicating that there is no clear trend.

\subsection{Candlestick Representations}\label{candRep}

While price movements in the stock market can be represented as a continuous curve that  is smoothed by time averaging over some period (be it seconds, minutes, hours, or days),
 candlesticks were proposed as a tool to better visualize the movements.  Candlesticks provide a summary of prices using four numbers - open, close, high, low - in a given period. In addition to the opening $O_t$ and closing $C_t$ values defined above, we now introduce:
 \begin{itemize}
 \item The highest value $H_t$  in a given interval is  the {\bf high}; 
 \item The lowest value $L_t$  in a given interval is  the {\bf low}.
 \end{itemize} Typical lengths of the periods that candlestick describe are one day, an hour, 30 minutes, and 5 minutes. 
Specifically, given a time series over a certain period, a candlestick $\mathcal{I}_t$ for the interval $t$ is defined by the quadruple
\begin{equation}
	 \mathcal{I}_t=(O_t, C_t, H_t, L_t).
\end{equation}\noindent
Taking the period to be a single day, this implies that $C_t-O_t$ is the change in value over the day. For $O_t>C_t$ this indicates a decrease in the value of the time series during the day, while $C_t>O_t$ implies an increase. Moreover, $C_{t-7}-O_t$ is the change in value over a given week.

\begin{figure}[h]
\centerline{
\includegraphics[width =0.45\textwidth]{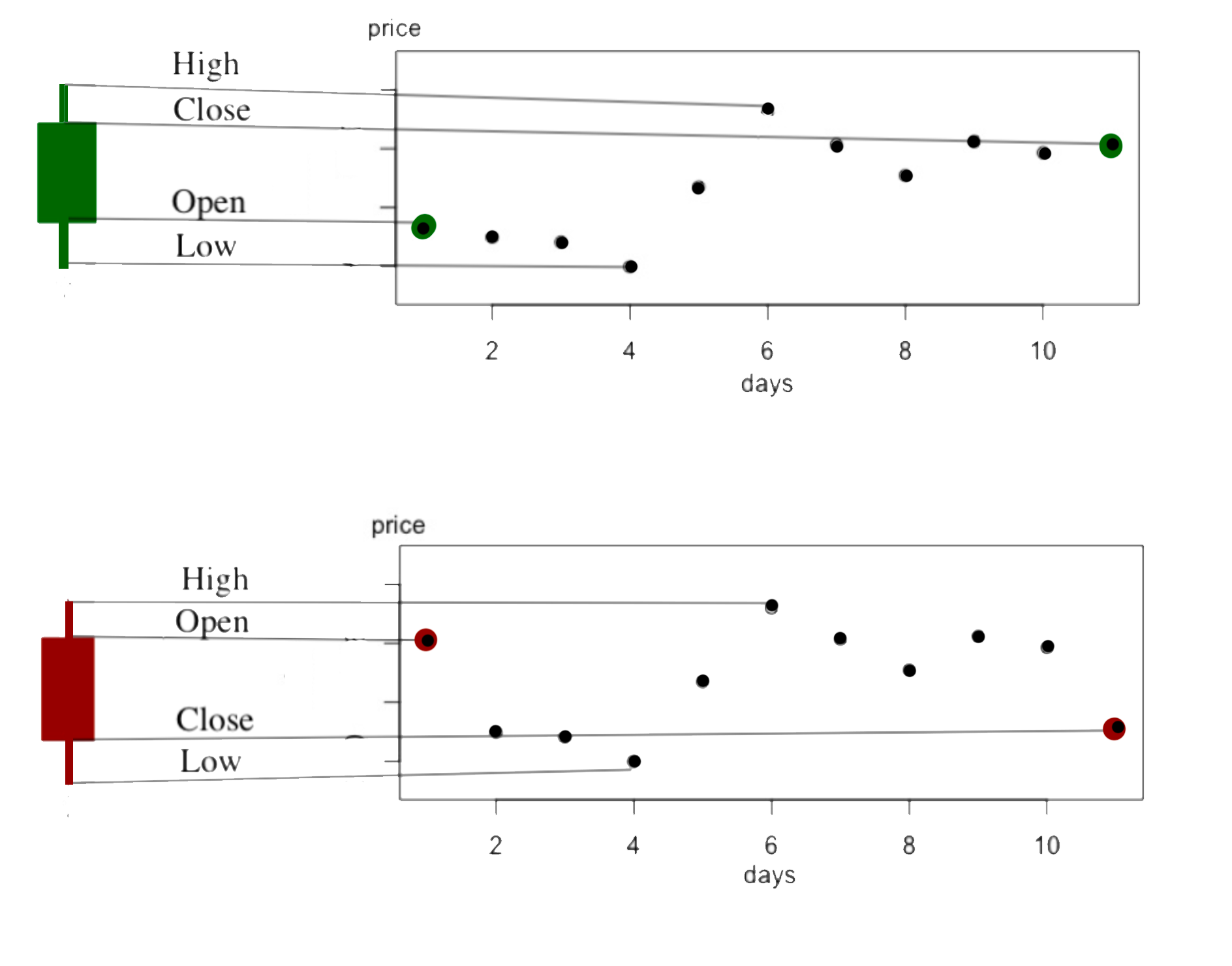} }
\caption{Illustrations of the construction of candlesticks. A red candlestick represents a decrease in value during the intervening period, observe that the open price is higher than the price at close. A green candlestick, conversely, indicates an increase in value. The proportions of the candlesticks are set by the open, high, low, close values over the  period.}
\label{Fig:defi}
\end{figure}

A visualization of how a single candlestick is constructed from the data in the intervening period is shown in  Figure \ref{Fig:defi}, following common practice we color the candlesticks red if $O_t>C_t$ and green if $C_t>O_t$, the color indicating whether the price increased or decreased over the period of the candlestick.

Each candlestick is comprised of three parts, the real body, and its lower and upper shadows.
The  real body $r_t$ at time $t$ is the difference between the opening values and the closing value
 \beq
 r_t = \text{abs}(O_t-C_t)~.
\eeq
This is represented as the central solid rectangle in the visualization of  Figure \ref{Fig:defi}.
  The lower shadows $l_t$ and upper shadows $u_t$ at time $t$ are defined by
\begin{eqnarray}
 	 l_t &=& \min(O_t, C_t)- L_t,\\[4pt]
 	 u_t &=& H_t- \max(O_t, C_t).
\end{eqnarray}
These are represented as the thin lines which extend above and below  the real body in the visualization of  Figure \ref{Fig:defi}. Note that in some cases the shadows may have vanishing extent, for instance for $O_t=L_t$ with $C_t< O_t$. 

In the context of the stock market, asset traders may often choose to utilize these discrete candlesticks  to visualize the data, as this representation provides substantially much information than the simpler line graphs of stock prices. 
Traders have developed a number of visual cues based on this candlestick representation --known as candlestick patterns-- which are thought to forecast future asset price moves, as we discuss next.

  \begin{figure}[h]
\centerline{
\includegraphics[scale=0.6]{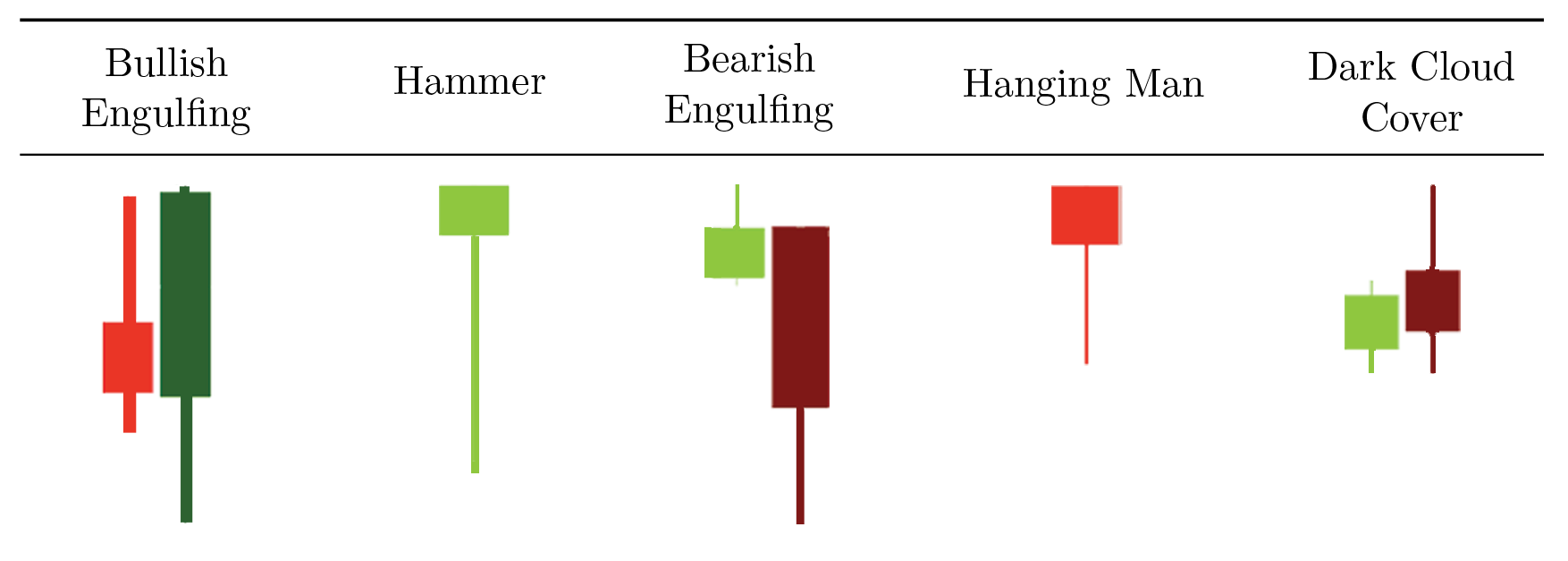} }
\caption{Visual definitions of five  candlestick patterns.}
\label{defi1}
\end{figure}

  \begin{figure*}[t!]
\centerline{
\includegraphics[scale=0.55]{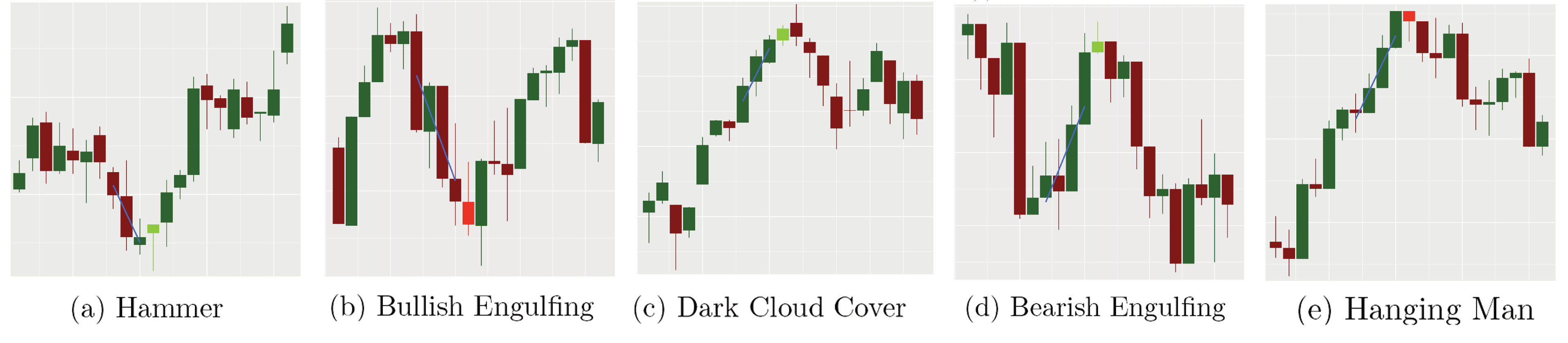} }
\caption{Examples of accurate forecasting via candlestick patterns. The $x$-axis provides an index of time with each candle representing one time period, while the $y$-axis indicates the value of some positive-valued measurable quantity (traditionally, share price). Axes values have been omitted as they are unimportant for these illustrations. The blue line indicates the 4-day trend lines established via linear regression, confirming either an appropriate uptrend or downtrend. The light colored candle indicates the start of each candlestick pattern, observe that in all cases shown the pattern corresponds to a trend reversal.}
\label{defi}
\end{figure*}

  \begin{figure}[b]
\centerline{
\includegraphics[scale=0.8]{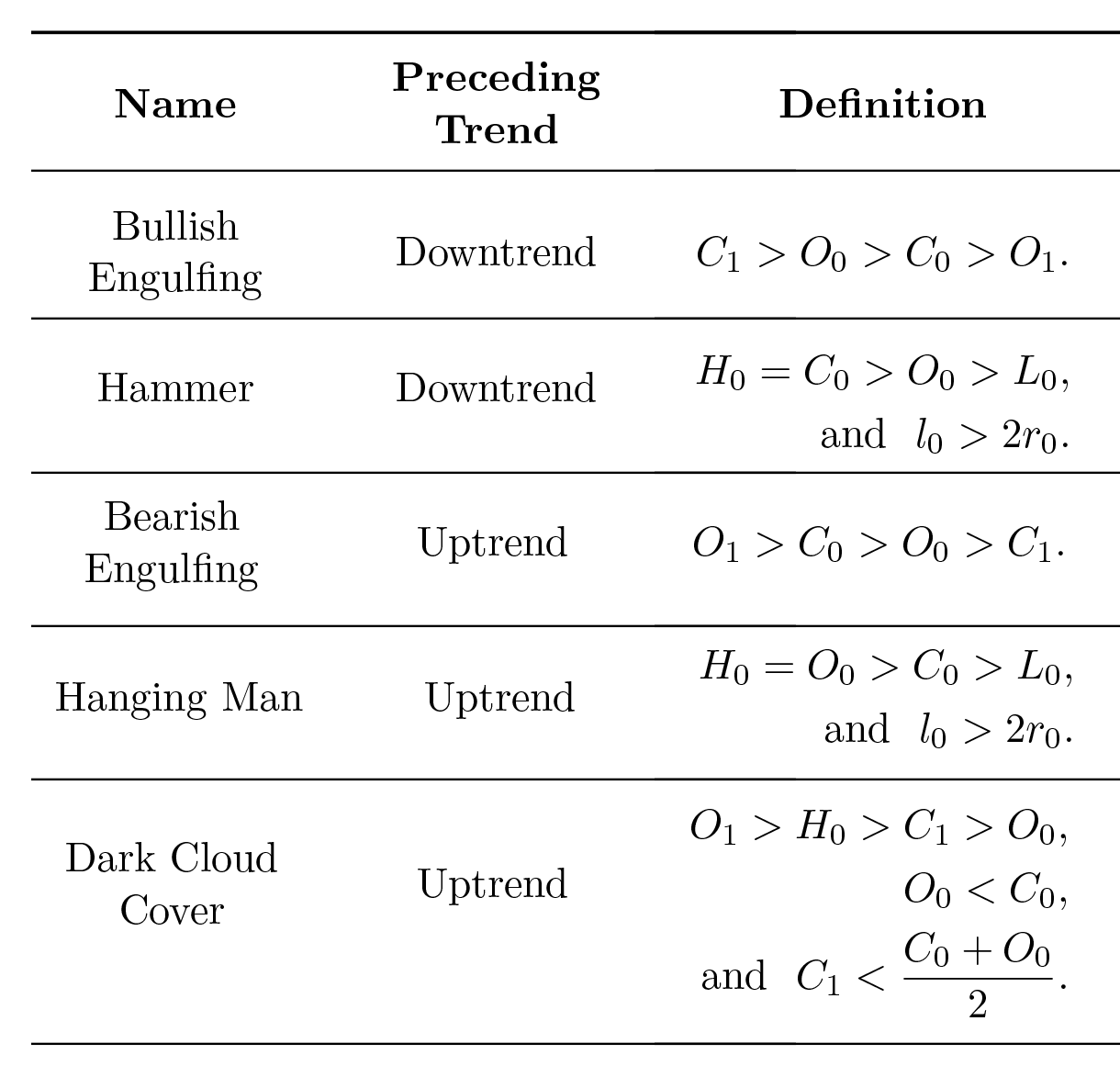} }
\caption{Mathematical definitions of five candlestick patterns.}
\label{defi2}
\end{figure}

\subsection{Candlestick Patterns}\label{cand_pat}
 
Candlestick patterns typically involve the relative magnitude of the high, low, open, and close values of one or two consecutive candlesticks. 
There is a widespread use of these patterns within the trading community, with the belief that specific configurations of candlesticks can be used to forecast future price movements \cite{bulkowski1}:
\begin{itemize}
\item  If a pattern predicts an uptrend will reverse to a downtrend, it is called a {\textbf{bearish reversal}} pattern. 
\item Conversely, a {\textbf{bullish reversal}} pattern predicts a reversal of a downwards trend to an uptrend. 
\end{itemize}

In this work we focus our analysis on three bearish reversal patterns patterns (Bearish Engulfing, Hanging Man, and Dark Cloud Over) and two bullish reversal patterns (Bullish Engulfing and Hammer). For the mathematical definition of the candlestick patterns we used the definitions proposed in \cite{profit_taiwan},  through restrictions on their $O_t, C_t,L_t, H_t$ values and requirements on a pre-existing trend. These patterns are shown graphically  in Figure \ref{defi1} and then are defined mathematically in Table \ref{defi}. 

The indices appearing in the definitions of  Table \ref{defi} denote the time ordering such that the first candle of each pattern occurs with time stamp $D=0$, with the second candle (if any) for $D=1$. We require a trend for the preceding $\delta$ intervals, such that there is an appropriate trend over the period $D-\delta\leq t \leq D$ as outlined in Section \ref{sec1.1}.

 Using R we implemented a code which takes time-series data and outputs candlestick representations then scans the output for specific patterns. Some example candlestick patterns identified by our code when applied to the  S\&P 500 Index (GSPC) daily data are presented in Figure \ref{defi}. We show the signal event which indicates the candlestick pattern  in a lighter shade. The regression line for the center of the four candlesticks preceding the candlestick patterns is shown to confirm the trend (note that we vary the required trend period in later  sections).

\subsection{Moving Average Convergence Divergence}\label{macd_sec}
The Moving Average Convergence Divergence (MACD) \cite{macd1} provides an alternative set of bullish/bearish market signals which can be repurposed for general time series data. The MACD is calculated using two exponential moving averages (EMAs),  calculated over two periods of differing length $n$. Specifically, for a given dataset of length $n$, usually the closing values $\{C_1,C_2, \cdots C_{n}\}$, the EMA $V_n$ is calculated recursively via 
\beq
\label{mm}
V_i [C_i]:= 
\begin{cases}
	C_1 &i =1\\ 
	s C_i + (1-s) V_{i-1} & i>1
\end{cases},
\eeq
 where   $s = \frac{2}{n+1}$ is smoothing factor.
 Thus, $V_n$ can be seen as  the exponential average over $n$ intervals, which by substitutions in the recursive formula can be expressed 
\beq
V_n = R[C_n + (1-R)C_{n-1} \cdots (1-R)^{n-1}C_1]~.
\eeq \noindent
Observe that the coefficient of each term decreases exponentially for earlier values in the time series, thus giving greater weighting to more recent data, hence the name. Given the EMA, the MACD is defined by the  difference between a longer period average $n_2$  and a shorter period average $n_1$ (thus by convention $n_1<n_2$), as follows
\beq
	\textrm{MACD}(n_1,n_2)= V_{n_1}- V_{n_2}~.
\eeq

\begin{figure}[t!]
\centering
\centerline{\includegraphics[width = 7.8cm]{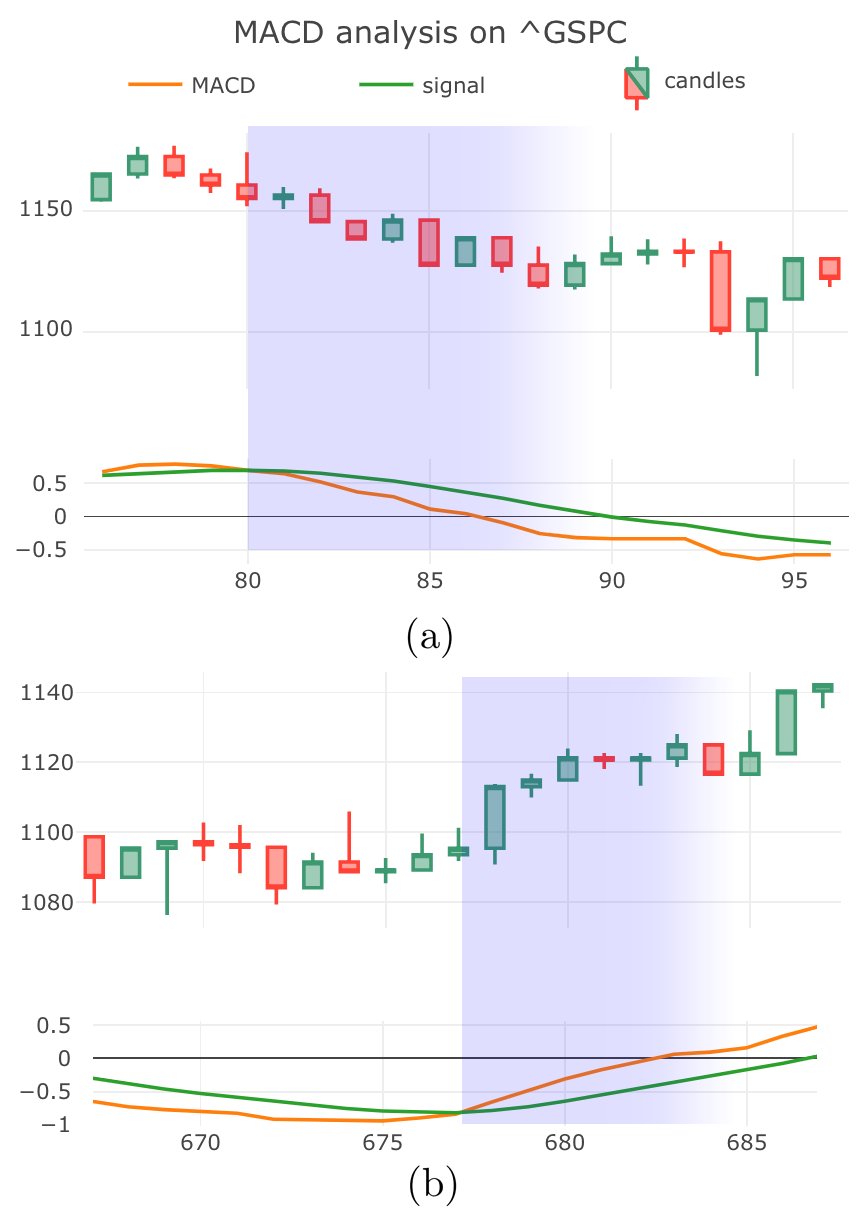}}
\caption{Examples of successful MACD signals, showing number of new COVID cases (y-axis) as a function of time in weeks from the first case. Panel (a) shows the MACD crossing the signal from line below, leading to bullish reversal.  Panel (b shows the converse bearish MACD signal.}
\label{Fig:macd}
\end{figure}

Common choices for $(n_1, n_2)$ are $(12,26)$, which corresponds to the number of trading days in roughly two weeks and a month, and lead to the following (Figure \ref{Fig:macd}):
\begin{itemize}
\item
When the MACD has large positive values, it indicates that the values have risen more in the recent $n_1$ observations when compared with the last $n_2$ observations, signifying a strong uptrend.
\item Conversely, when MACD is negative, the price has fallen more in the last $n_1$ observations, signifying a recent downtrend.
\end{itemize}

MACD analyses provides signals based on ``momentum'' of the time series. To identify buy and sell signals, the MACD is compared to the so-called Signal Line $S$, defined by 
\beq
S = V_{n_3}[\textrm{MACD}(n_1,n_2)].
\eeq
A common value for $n_3$ is $9$, signifying a week and a half trading period. 

There are many ways to use the signal line. In this paper we will focus on crossovers between the MACD and $S$,  illustrated  in  Figure \ref{Fig:macd}, which are described below:

\begin{itemize}
\item When MACD crosses from below to above the signal line, it serves as a bullish signal because the crossing signifies a strong uptrend in MACD, meaning the short-term momentum has risen faster than the long term momentum. 
\item Conversely, when MACD crosses from above to below the signal line, it serves as a bearish signal forecasting a downturn in values.\end{itemize}

 \begin{figure}[t!]
\centerline{
\includegraphics[width = 8cm]{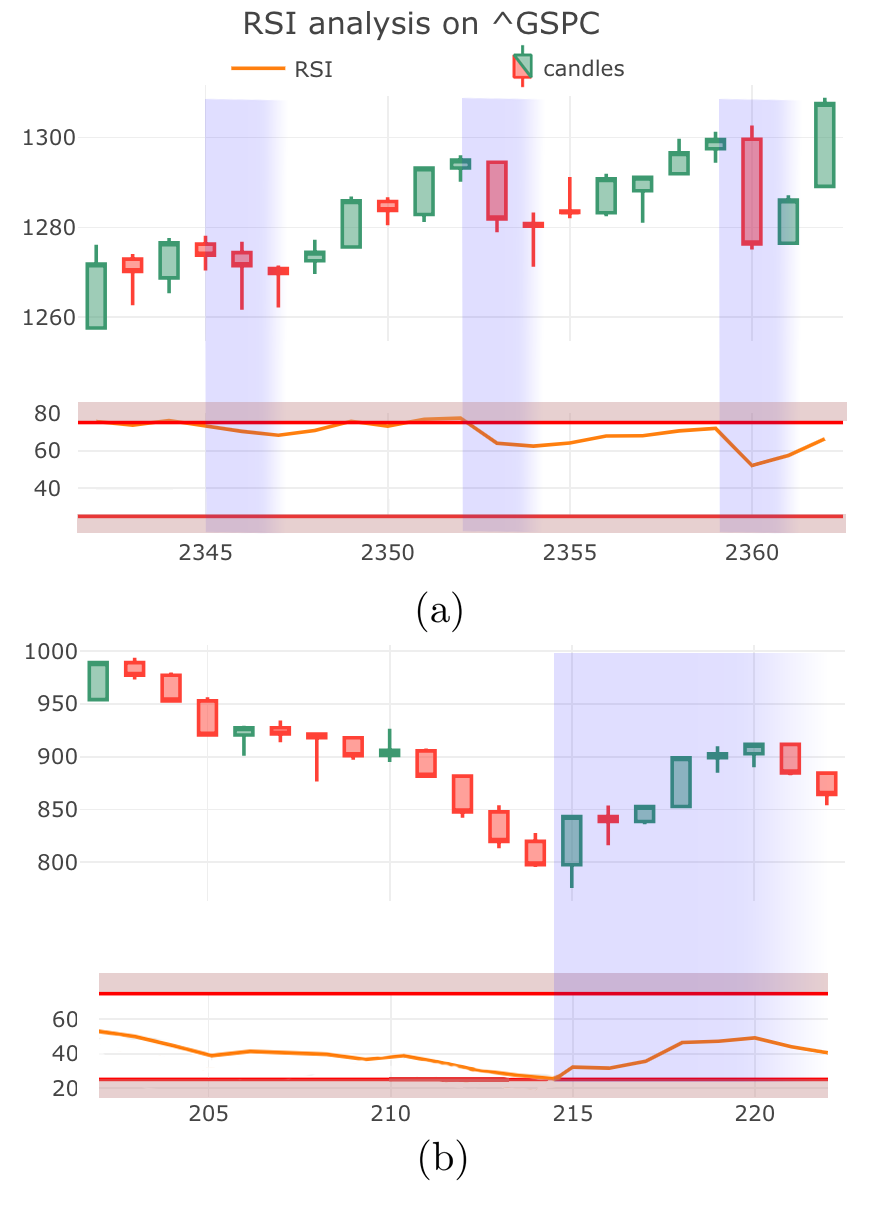}}
\caption{Examples of successful  RSI signals. a).~RSI index is low, predicting reversal to an uptrend as observed. b).~RSI index is high, signally the start of a downtrend.}
\label{Fig:rsi}
\end{figure}

 \subsection{Relative Strength Index}\label{RSI_sec}
The Relative Strength Index (RSI)  quantifies the momentum in the times series data through average rate of increases and decreases in value (see Figure \ref{Fig:rsi}). The indicator is constructed by dividing the closing values $\{C_t\}$ over some period into two sets: 
\begin{itemize}
\item The set $\{G_t\}$ in which the series increased:
\begin{eqnarray}\{G_t\} := \frac{C_t-C_{t-1}}{C_t}, \qquad C_t>C_{t-1}.
\end{eqnarray}

\item  The set $\{D_t\}$  in which the series decreased:
\begin{eqnarray}\{D_t\}:= \frac{C_{t-1}-C_t}{C_t}, \qquad C_t<C_{t-1}.
\end{eqnarray}
\end{itemize}

From the above sets one can compute the averages $\bar G_t$ and $\bar L_t $ using the EMA $V_n$ over $n$ periods with a smoothing factor $s = \frac{1}{n}$, leading to
\beq
	\bar G_t = V_n[G_t]~~~{\rm and}~ ~~D_t =  V_n[D_t]~.
\eeq
Then,  the RSI$_t$ indicator at time $t$ is defined as follows
\beq 
\textrm{RSI}_t := 100- \frac{100}{1+\frac{\bar G_t}{\bar D_t}}~.
\eeq

In the stock market the RSI is  used to signal when an asset has become overbought (meaning it has appreciated more rapidly than thought to be typically sustaiable) or oversold. In particular, a high RSI is thought to indicate that one should anticipate a reversal from an uptrend to a downtrend in the near-term. A low RSI is interpreted in by market traders as an asset being oversold, and predicts a near-term increase in prices.  We set the threshold for high and low RSI$_t$ to be $75$ and $25$. When the RSI reaches $25$ it serves as a bullish signal and conversely, when the RSI reaches $75$ this gives a bearish signal.  Figure \ref{Fig:rsi} gives two examples of accurate RSI signals.

\section{Statistical Methods}
\label{Sec3}

Following a standard hypothesis testing protocol, a given procedure tests a {\it null hypothesis} $H_0$, against a {\it alternative hypothesis} $H_1$. The testing framework then either rejects, or fails to reject, the null hypothesis. Specifically, to test whether or not these technical indicators can correctly predict trends, we formulate a testing procedure using the Wilcoxon Signed-Rank Test. 

\vspace{-1mm}
\subsection{Wilcoxon Signed-Rank Test}
\vspace{-1mm}

The Wilcoxon test \cite{wilcox1} is  a nonparametric test for testing the median of a distribution, as we outline below. Previous studies by Goo,~Chen, \& Chang \cite{profit_taiwan} utilized the t-test as a possible way to confirm the predictive powers of candlestick patterns. However, the t-test is a parametric test, meaning that it assumes the normality of the observations. The t-test studies the means of the given sample which only makes it reliable in normal samples. Notably, when \cite{profit_taiwan} was published (2007), normal distributions were a common belief for the rate of return in stock markets, however, recent studies have suggested that Laplace distributions fit better to the daily return of stock markets \cite{laplace}. Given that the distribution of an $n$ day return cannot be assumed as normally distributed, a non-parametric test that tests for the median, i.e.~the 50th percentile of the distribution, is the most desirable. 

Specifically, here we employ a One-Sample Wilcoxon Test to test for a hypothesized median.  Suppose that at time $t$ one observes a signal from one of the technical indicators outlined in the previous section, i.e.~a candlestick pattern, a high/low RSI$_t$ reading, or a crossover in the MACD. We record the value of the close of that day and denote it as $C_1$, we also record the closes for $n$ days following the signal and form a vector of values $\vec C = \{C_1, C_2, \ldots, C_n\} $. From this we can calculate the rate of return $R_i$ for $i$ days after the observation of the signal event as follows
\begin{equation}
	R_i = \frac{C_{i+t}- C_{t}}{C_{t}}~.
\end{equation}
Thus we can also define a vector of rates of return following the signal event: $\vec R = \{R_1, R_2, \ldots, R_n\}$.

To proceed, we take the $n$-day rate of returns vector $\vec R = \{R_1, R_2, \ldots, R_n\}$ and denote the median of the set by $\tilde R$. Then we define
$\{d_1, d_2, \cdots, d_n\}$ to be the difference of each $R_n$ from the median $\tilde R$ such that
\begin{equation}
d_t = (R_t-\tilde R)~.
\end{equation}
The null hypothesis $H_0$ and  alternative hypothesis $H_1$ for the pooled sample of the rate of returns on a given day $\vec R$ can be formulated as follows. Hypothesis $H_1$ holds that at the occurrence of a bullish (bearish) signal event there should be a positive (negative) rate of return in the near-term future $\tilde R > 0$ (conversely, $\tilde R < 0$). The null hypothesis $H_0$ holds that the rate of return should be uncorrelated to the signal events, implying that the indicator under examination fails to provide accurate forecasts.

To implement the one-sample Wilcoxon test we assign to each $d_t$ sequential integers $R(d_T)\in\mathbb{Z}$ (a rank), assigning $R=1$ to the $d_t$ with the smallest absolute value, $R=2$ for the next smallest absolute value, and so forth, such that the rank of the $d_t$ with the largest absolute value is $R=n$. Then we define $W_1$ and $W_2$ as follows
\beq
	W_1 = \displaystyle \sum_{d_t>0} R(d_t),~~~{\rm and}~~~ 
	W_2 = \displaystyle \sum_{d_t<0} R(d_t)~.
\eeq
The Wilcoxon test statistic $W'$ is then defined to be 
\begin{equation}
	W'= W_1-W_2~.
\end{equation}

Note that the statistic $W'$ is essentially the difference between all the ranks of the observations below the hypothesized median ($W_1$) and the ranks of the observations above the hypothesized median ($W_2$). It is a robust way to measure a median since, if the true median is the hypothesized median,  the distribution of samples should be symmetric about the median. Thus, when we rank the difference of the samples from the median, about half should be positive and about half should be negative, and the sum of their ranks should cancel. 

The distribution of expected outcomes assuming $H_0$ is true, the null distribution, is centered on $\tilde R = 0$. In contrast, suppose that $H_1$ is true,  then for a bullish reversal $\tilde R >0$ and there should be fewer observations below $0$ (the null hypothesis median).  In this case the true distribution is shifted toward the new median $\tilde R >0$ and,  as a result, the test statistic $W'$ would tend to be greater.

If the true median is sufficiently different from zero, we reject the claim that its median is $0$. This statement can be reformulated in terms of a random variable following the null distribution $W$, such that for bullish reversals the null hypothesis $H_0$ is rejected for
\beq\label{PWW}
P(W> W')=\int_{W'}^\infty p(t)~{\rm dt} \leq \alpha,
\eeq
where $p(t)$ is the Wilcoxon distribution and $\alpha$ is a constant threshold which signifies thee significance level, commonly $A$ is taken to be $0.05$ or $0.10$. Conversely, for bearish reversals the $H_0$ is rejected for $P(W< W') \leq \alpha$.

\newpage
\subsection{Calculating $p$-values}

Given the above we can compute the  $p$-value from eq.~(\ref{PWW}) to quantify the statistical significance of  positive correlations. The $p$-value is the probability that the obtained statistic (or a more extreme statistic)  occurs given that the null hypothesis is true.   A low $p$-value, below the significance levels $\alpha$, signifies the unlikeliness of the null hypothesis being true. Typical values for $\alpha$ are $0.05$ and $0.10$, we will adopt the latter value going forward.  For an observed $p$-value $p$ such that  $p\leq\alpha=0.1$, we reject the null hypothesis and claim that the alternative hypothesis is more likely. In our testing procedure, a rejection of the null hypothesis would indicate that a given technical indicator has predicting power.

We wish to quantify whether a signal event identified by one of the technical indicators discussed above makes predictions which are statistically, significant. Specifically, the signal events under consideration are
\begin{itemize}
\item Occurrences of a Candlestick Patterns;
\item MACD cross over events;
\item RSI values of 25 or 75. 
\end{itemize}
Out R code scans the data for such signal events and then record each occurrence along with the $O_t,C_t,H_t,L_t$ values for a range of days or weeks around each occurrence. We use the first $\delta$ time intervals prior to the signal to establish the initial trend, the value of $\delta$ is stated in the analysis. We then take the $\Delta t$ time intervals after the signal events to assess whether the signal correctly predicted the future evolution of the time series. 

 As an initial approach towards quantifying the statistically significance of the predictions we took the set of signals and calculated the $p$-value multiple times for different choices of $\Delta t$. This analysis is intuitive and insightful (and we present results arising from this in the Appendix), however, there are a number of critical issues:
 \begin{itemize}
\item It gives multiple $p$-values for each indicator (which can be conflicting).
\item The $p$-values for each $\Delta t$ are not independent.
\item For 9 indicators and 10 choices of $\Delta t$ one calculates 90 $p$-values, thus one anticipates false positives.
 \end{itemize}

Hence, below we outline a more sophisticated analysis. We start with the record of all occurrences of signals for a given indicators identified by our code. Then to calculate a global $p$-value for a given indicator, we subdivide the data of corresponding to each occurrences randomly into $n$ subsets. Given that each signals typically occur $\mathcal{O}(100)$ times in the data series (cf.~Table \ref{summary_covid}), we will use $n=3$ subsets. Then for each subset we calculate the $p$-value using a different $\Delta t$ for each subset. 

Notably, these subsets will be independent of each other, allowing us to calculate independent $p$-values for each subset. Since these $p$-values are independent we can then use the standard Fisher method \cite{Fisher} to combine the three $p$-values $p_i$ (with $i=1,2,3$) into the following statistic 
\beq
X=-2{\rm log}(p_1)-2{\rm log}(p_2)-2{\rm log}(p_3).
\eeq
The  statistic $X$  has a chi-squared distribution with 6 degrees of freedom (more generally $2n$ for $n$ subdivisions of the dataset).
To obtain the global $p$-value for each indicator we then calculate the area under the chi-squared curve (with 6 degrees of freedom) which lies to the right of the value of $X$. Following this procedure, we report our finding for the case of new COVID infections in Table~\ref{results_covid} and in the context of the stock market in Table~\ref{summary_market}.

\section{Predicting  New Cases of COVID-19}
\label{Sec4}

With the technical indicators defined in Section \ref{Sec2} and the statistical analysis of Section \ref{Sec3}, we are now prepared to examine whether technical analysis can provide statistically significant predictions when adapted to study the near-term changes in the number of new COVID-19 daily cases. Following this we will outline and evaluate two specific use cases for these indicators, namely, identifying the peak of a wave of infections, and the commencement of subsequent pandemic waves.

\subsection{Statistical Significance}
\label{Sec4A}

To investigate whether these technical indicators could be of use during a pandemic, we undertook an analysis of World Health Organisation (WHO) COVID-19 data. Specifically, we used data on the daily reported cases for $237$ countries from January, 3, 2020 to July, 29, 2021 \cite{covid}.
 For a given country, the starting date of the pandemic was defined to be the identification of the first case. 
 
 We then grouped the observations of new cases into weekly candlesticks $\{P_c\}$ by setting the open values as the number of new  cases on the first day of each 7-day period and the close values as the number of new cases on the last day of each 7-day period. The real body of each candle was defined using the highest/lowest number of new daily cases during the corresponding 7-day period. 
The data was organized into $17140$ candlesticks with about $70$ candlesticks for each country. 

Our code identified occurrences of the various signals relating to the technical indicators under consideration across all countries, and we present the number of observation for each signal in  Table \ref{summary_covid}.  For calculating of the pre-existing trends to identify candlestick patterns, we use the two preceding weekly candlesticks. 
Following Section \ref{Sec3}, after identifying the occurrences of each indicator in the COVID datasets for each country, we pooled the occurrences together to proceed with the statistical analysis. We dropped any indicator with less than 50 occurrences from our analysis.

\begin{table}[b]
\centering 
\begin{tabular}{c|c}
	\text{Signal Name} & \text{Number of observations} \\
	\hline
	\text{Bullish Engulfing} &  99\\
	\text{Bearish Engulfing} & 123\\
	\text{Hammer} & 127\\
	\text{Hanging Man} &  156\\
	\text{Dark Cloud Over} & 30\\
	\text{Bullish MACD} & 217\\
	\text{Bearish MACD} & 245\\
	\text{Bullish RSI} & 46\\
	\text{Bearish RSI} & 1057\\
\end{tabular}
	\caption{Summary of the number of observations of technical analysis signals in pooled 17140 weekly candlesticks of COVID-19 data for $237$ countries from $\textrm{January},3,2020$ to $\textrm{July} ,29, 2021$ data obtained from WHO \cite{covid}.
	\vspace*{2.1mm}}
	\label{summary_covid}
\end{table}

\begin{table}[b]
\centering 
\begin{tabular}{c|c|c}
	\text{Signal Name} &~ \text{$p$-value}  ~& ~~Significance \\
	\hline
	\text{Bullish Engulfing} &  $\bf 0.0005 $&  3.2$\sigma$ \\
	 \text{Bearish Engulfing}  & 0.63 & -- \\
	 	\text{Hammer} & $\bf 0.014$ & 2.2$\sigma$  \\
	\text{Hanging Man} &  0.63 & -- \\
\text{Bullish MACD}	 & $\bf 0.0071$ & 2.5$\sigma$ \\
\text{Bearish MACD}	& $\bf 6.8\times10^{-9}$ &  5.7$\sigma$ \\
\text{Bearish RSI}	 &0.63 & -- \\
\end{tabular}
	\caption{Statistical significance of each technical indicator is shown in terms of their global $p$-value (averaging over multiple values of $\Delta t$). Statistically significant $p$-values (those $<0.1$) are shown in bold, and for those indicators we also give the significance in terms of their $\sigma$ (equivalent to a $Z$-score).  
	\label{results_covid}
}
\end{table}

 Since we only have daily data (and not hourly) we choose to construct weekly candlesticks.  For the analysis of all of the indicators we calculated the $p$-value on a weekly time scale, thus taking $\Delta t$ to be some $\mathcal{O}(1)$ number weeks following a signal observation. 
We subdivided our data as described in Section \ref{Sec3} and calculated the global $p$-value for each indicator by averaging over three choices for $\Delta t$, specifically we used   $\Delta t=3,5,7$ weeks. The combination of the individual $p$-values is described in  Section \ref{Sec3}. We present the global $p$-values from our analysis in Table~\ref{results_covid}. Additionally, as a preliminary analysis we calculated the $p$-value while varying $\Delta t$ to see the impact, this analysis is presented in the Appendix (however, as alluded to in Section \ref{Sec3}, while this can be insightful, it encounters some technical drawbacks). 

By observation of Table  \ref{results_covid} it can been seen that some indicators certainly are statistically significant predictors of future COVID cases, while others are not. Specifically, 
Bullish Engulfing and (bullish) Hammer candlestick patterns, as well as both MACD indicators are all seen to be  statistically significant. Notably, the $p$-value of the Bearish MACD signal implies that this is a highly accurate indicator. Note that the Dark Cloud Over and Bullish RSI indicators  had only $\mathcal{O}(10)$ occurrences, this is a relatively small sample which could lead to an erroneous conclusion, and thus were omitted from our analysis. We also highlight that these global $p$-values are corroborated by the cruder --although perhaps more intuitive-- local $p$-value analysis in the Appendix.

Having concluded that a subset of technical indicators can indeed provide insights into the near-term progression of the pandemic, we next explore how these indicators might be applied to gain insights into trends during an ongoing pandemic. Specifically, we next explore two particularly important use cases:
\begin{itemize}
\item Identification of the peak of a pandemic waves.
\item Forecast the start of a new wave of a pandemic.
\end{itemize}

\begin{figure*}[t]
\centerline{
        \includegraphics[height=33mm]{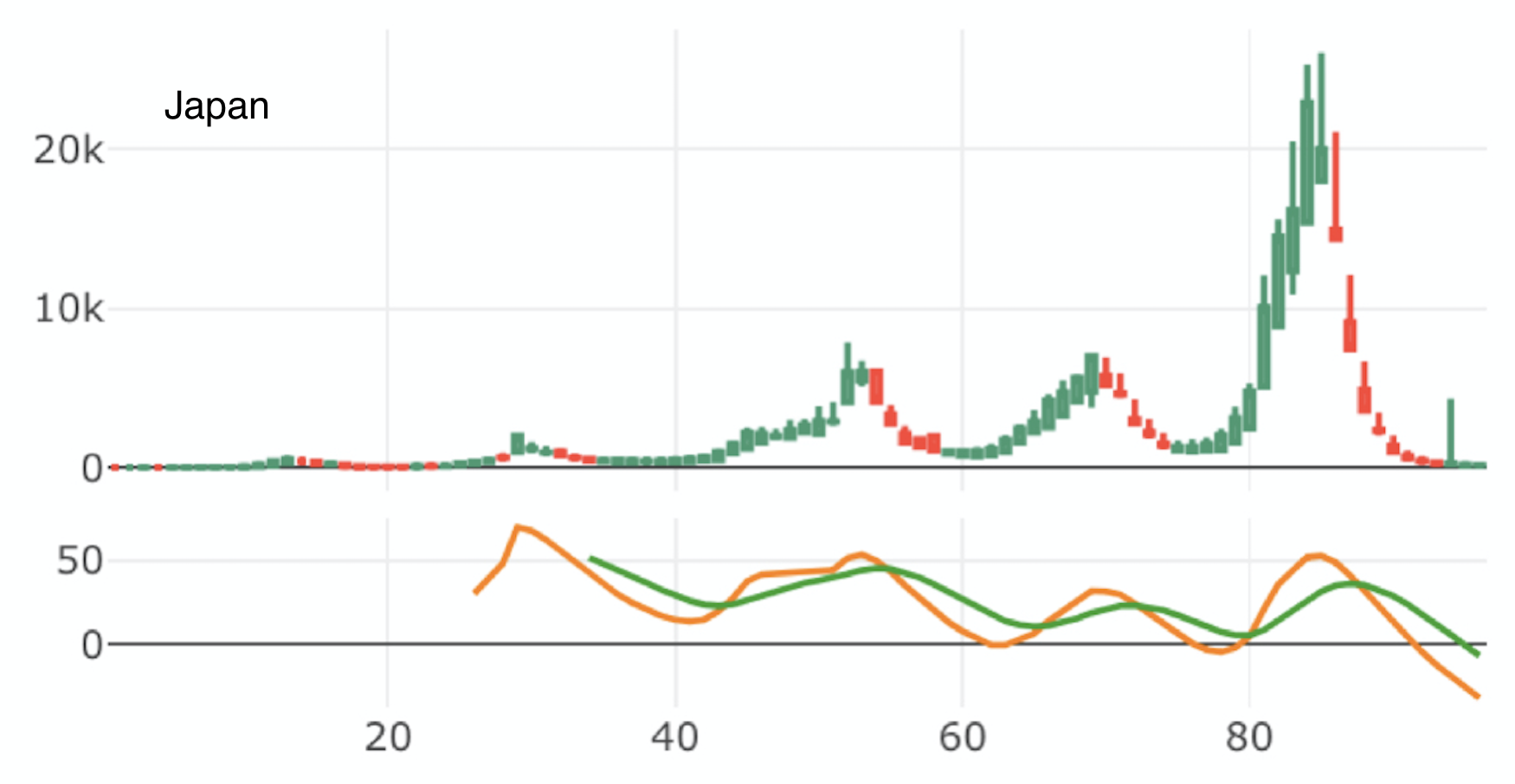} \hspace{1mm}
        \includegraphics[height=33mm]{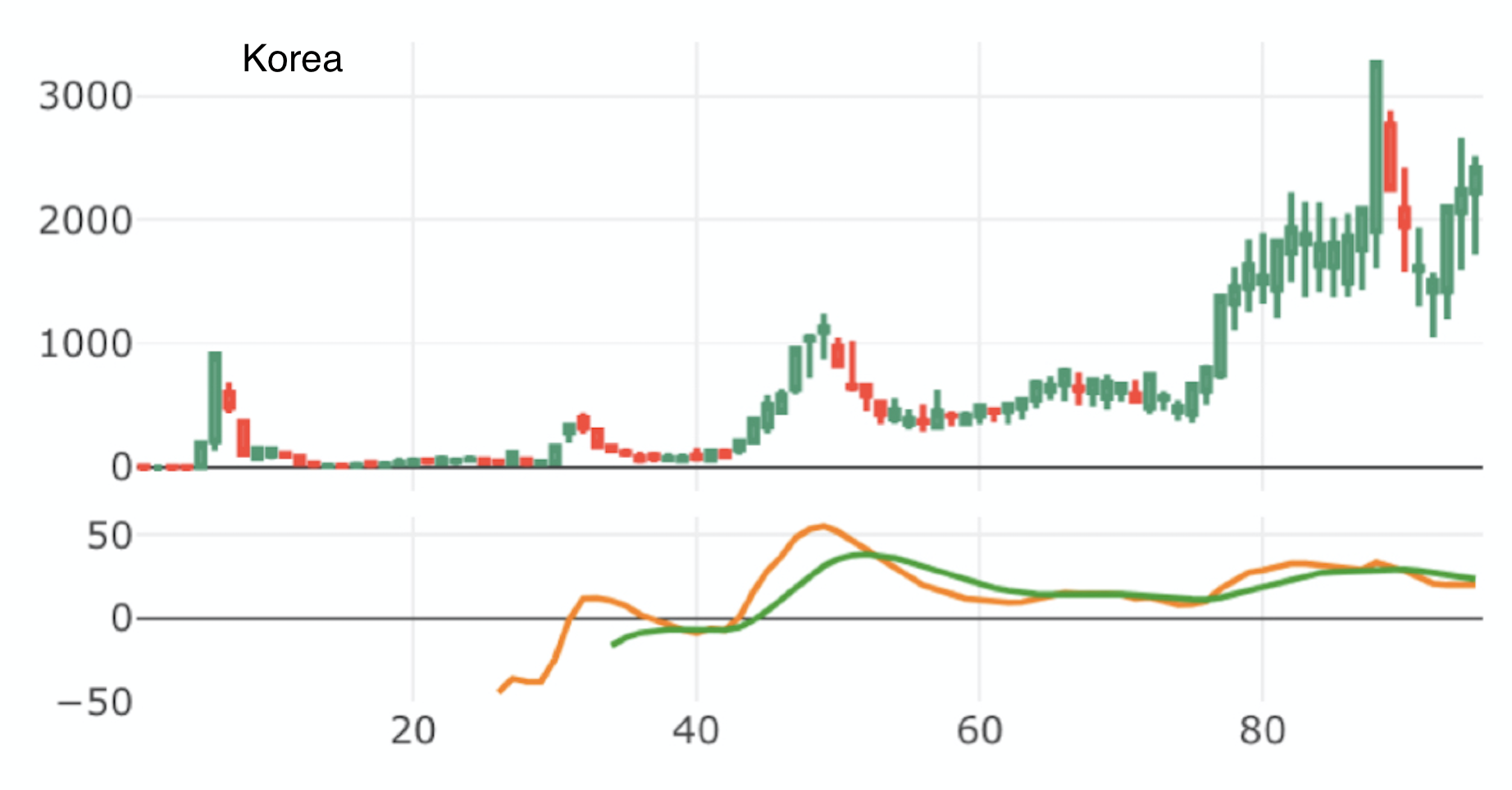}  \hspace{1mm}
        \includegraphics[height=33mm]{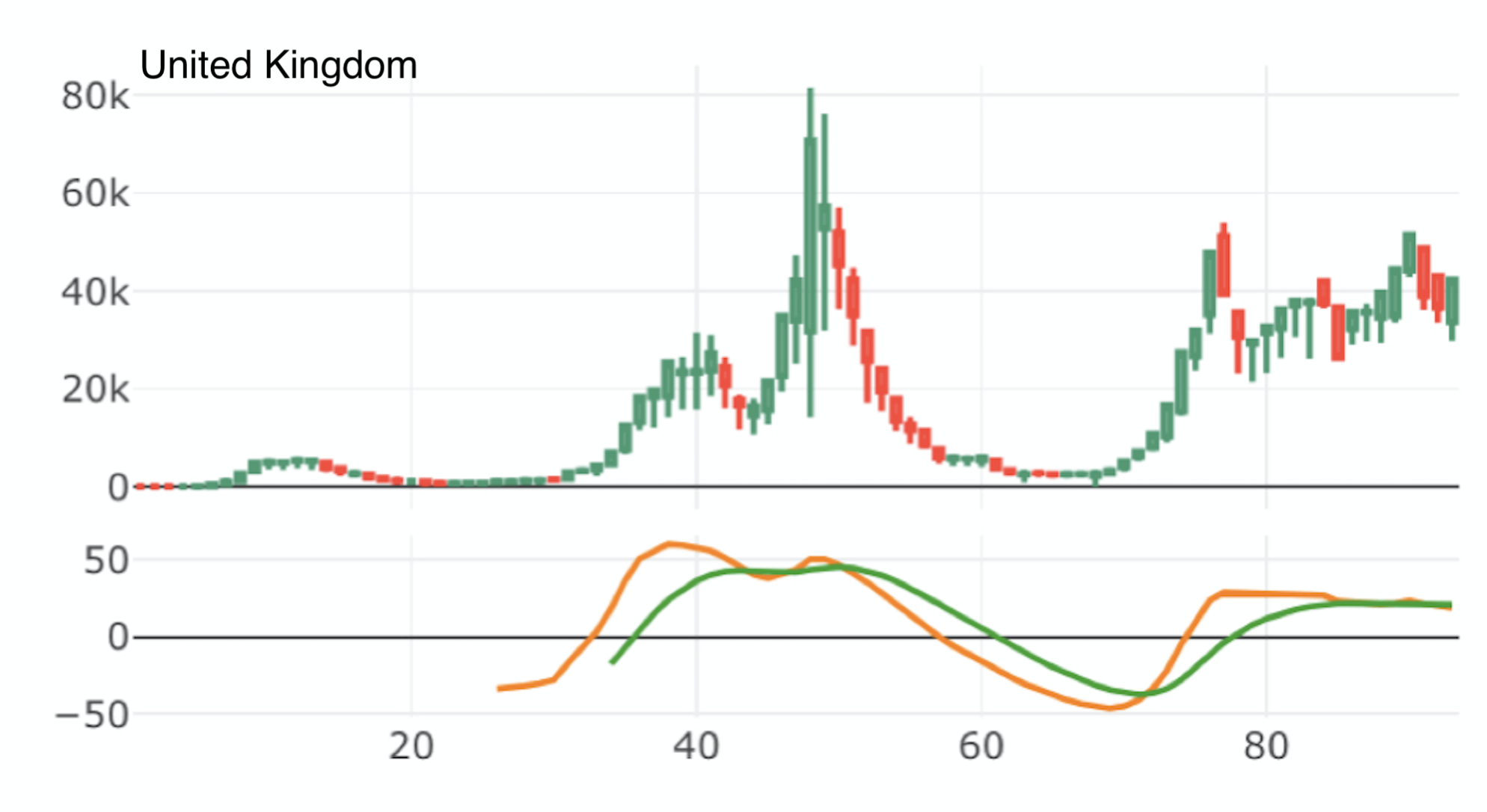} }
    \caption{
Candlestick plots of new COVID cases in Japan (first case: 2020/01/14), 
Korea (first case: 2020/01/19)
and the UK (first case: 2020/02/01). Below each plot is the associated MACD chart. A bullish MACD cross-over involves the (red) MACD curve crossing the (green) Signal curve from below, this predicts a rise in cases in the near-term. These examples  support that the MACD analysis can provide useful information regarding the evolution of the pandemic.
    \label{lastfig}}
\end{figure*}

\subsection{Peaks of Pandemic Waves}
\label{Sec4B}

In the early stages of a pandemic (such as the current COVID-19 crisis) new daily cases grow steadily from week-to-week, perhaps with some small daily fluctuations. At this stage the number of COVID cases exhibits a clear uptrend. Accordingly governments and health official put in place policies and funding to endeavour to reduce the spread of infections. A major milestone in controlling the pandemic is to identify a peak in the daily cases. While peaks are simple to identify in retrospect, at the height of a pandemic it is far from obvious whether a decline in cases is a fluctuation or a local top. The signals we have considered each imply reversals in the trend, thus if COVID cases are growing and one observes a bearish signal using weekly data, this is a prediction that cases will begin falling over the next few weeks. 

Therefore, the peak in the number of cases corresponds to a change in the trend of the pandemic, and this is precisely what the indicators that we have been studying are designed to detect. We now apply the Bearish MACD indicator in an effort to identify the peaks of the COVID-19 pandemic in a number of case studies. Notably,  the Bearish MACD was the sole indicators for bearish reversals that we found to be statistically significant in Table \ref{results_covid}. In other words, we propose that a crossing of the MACD line below the signal lines indicate a peak in infections. More specifically, this indicates the end of the first wave and one expects such crossing events to coincide with each of the peaks of the pandemic. 

In Figure \ref{lastfig} we present COVID-19 daily cases for Japan, South Africa, and the UK in the form of a weekly candlestick charts along with the corresponding plots of the MACD indicator. Observe that a change from uptrend to downtrend does indeed coincide with the MACD line crossing the signal line from above. Moreover, one can interpret the convergence of the MACD and signal lines as indication that cases are nearing a peak, which is indicative that current health policies are likely being effective in reducing the rate of infections.

	\vspace{-3mm}
\subsection{Additional Waves of the Pandemic}
\label{Sec4C}
\vspace{-3mm}

\begin{table}[b!]
\centering 
\hspace*{-5mm}\begin{tabular}{c|c|c}
	\text{Signal name} &$\#$ \text{Occurrences} (daily) & $\#$ \text{Occurrences} (weekly) \\
	\hline
	\text{Bullish Engulfing} &  290 & 67 \\
	\text{Bearish Engulfing} & 451& 104\\
	\text{Hammer} & 21& 31\\
	\text{Hanging Man} &  45& 129\\
	\text{Dark Cloud Over} & 82& 0 \\
	\text{Bullish MACD} & 4587 & 660\\
	\text{Bearish MACD} & 4584& 660\\
	\text{Bullish RSI} & 1133& 0\\
	\text{Bearish RSI} & 3588& 251\\
\end{tabular}
	\vspace{-3mm}
	\caption{Summary of the number of observations of technical analysis signals in pooled 120,000 sample of 28 stocks over 10 years used in our study from Yahoo Finance \cite{yahoo}.}
	\label{summary}
\end{table}

As evident from the COVID-19 crisis, pandemics can exhibit multiple waves of infections. A second wave refers to the case in which after an initial peak in infections there is a period in which new cases are in decline, then a subsequent reversal with daily cases growing once again. 

By inspection of  Figure \ref{lastfig} we can see that the transition from declining cases to increasing new cases can be  discerned through the observation of the Bullish  MACD  signal. We know that this is statistically significant predictor of new cases  and this is supported by the case studies of Figure \ref{lastfig}, where one can clearly see that there is an apparent correlation between the crossing event and the commencement of a second wave. The observation of a Bullish Engulfing or Hammer pattern in the candlesticks would also be indicative of subsequent waves. 

These tools have significant value for predicting the broad strokes of the future course of the pandemic and can used to identify when a relaxation of health restrictions  (such as ending social distancing or mask mandates) is leading to a new wave of infections. The MACD analysis for Japan (Figure \ref{lastfig}, left) is a particular good example of how this indicator gives clear signals of subsequent peaks. The observation of a bullish signal should be used as an indicator that health restrictions must be re-established in order to regain a downtrend in new cases.

\newpage
Since Figure \ref{lastfig} includes data up to November 2021, this provides a COVID-19 forecast for Winter 2021. For instance this suggests that a new wave of infections is not imminent for Japan, whereas since the Korean MACD is nearly crossing one anticipates there could be a growth in new cases in early 2022. The proximity of the MACD and signal lines in the UK plot is ambiguous and thus the near-term future is unclear, but this should be taken as an indicator that health restrictions should be strengthened to mitigate the risk of increasing infections.

\vspace{-3mm}
\section{Application in Stock Markets}
\label{Sec5}
\vspace{-3mm}
 
Finally, we also apply our statistical methods to a pool of stock market data. While other such studies have been undertaken, we believe that our use of the Wilcoxon signed-rank test and carefully averaging over multiple time periods in calculating the $p$-value make our analysis more robust.
Following the methodology of Section \ref{Sec3}, we applied our code on a pool of stock market data based on 28 stocks and indices, including companies such as Google, Amazon, indexes such as $S\&P $ 500. The data was all sampled with daily and weekly candlesticks obtained from Yahoo Finance \cite{yahoo}. Table \ref{summary} summarizes the number of observations for the aforementioned signals in the compiled data set. We dropped any indicator with less than 50 occurrences from our analysis.

 \begin{table}[b!]
\centering 
\begin{tabular}{c|c|c}
	\text{Signal Name} &~  $p$-value (daily) ~& ~~$p$-value (weekly) \\
	\hline
	\text{Bullish Engulfing} & {\bf  0.092}  &  0.41 \\
	\text{Bearish Engulfing} & 0.94 &  0.97  \\
	\text{Hanging Man} &  --  & 0.99 \\
	\text{Dark Cloud Over} &  0.26 &  --  \\
	\text{Bullish MACD} &   $\bf 1.4\times10^{-9}$ & {\bf 0.058} \\
	\text{Bearish MACD} & {\bf  0.0015}  & 0.19 \\
	\text{Bullish RSI} & {\bf  0.0015}  &  -- \\
	\text{Bearish RSI} & {\bf  0.1}  & 0.997
\end{tabular}
\vspace*{-1.1mm}
	\caption{Global statistical significant of each technical indicator for stock market data. We calculate the $p$-values for both weekly and daily partitions of the data. Statistically significant $p$-values (those $<0.1$) are shown in boldface.
	\label{summary_market}
}
\end{table}

\newpage
Table \ref{summary_market}  gives the $p$-value of each signal using the one-sample Wilcoxon signed-rank test.
 While for the COVID study $\Delta t$ was measured in weeks, as we have much more data for stock prices we undertake our analysis at the time scale of both weeks and days. We take statistically significance to be $p$-values less than $0.10$.

Our results indicate that for financial data the MACD and RSI signals, as well as the Bullish Engulfing pattern, are statistically significant on the daily timeframe. Although only the Bullish MACD signal is found to be significant for  financial data analysed on the weekly timeframe. We note that our findings regarding which candlestick patterns are predictive, disagree with \cite{profit_taiwan} which used a different methodology for their analysis.

\vspace{-4mm}
\section{Concluding Remarks}
\label{Sec6}
\vspace{-3mm}

The world has been struggling with the COVID-19 pandemic over the past two years, and increasing attention has been given to the forecasting of infectious diseases.  This paper has demonstrated that technical analysis  used in asset trading can be repurposed to forecast near-term changes in the number of new cases of COVID-19 and, more generally, in future pandemics. In particular, we analysed WHO data of the daily new COVID-19  cases for all countries and identified a number of   technical indicators that make statistically significant predictions. 

 Since financial data and COVID data arises from very different systems, it is notable that technical analysis can provide predictions in both settings. We conjectured that these indicators work across these two systems since both can be modeled as non-stationary random walks. It is conceivable that these indicators can identify underlying trends in the time series.   Moreover, some groups have expressed doubts regarding whether  technical analysis has any intrinsic predictive power, and these results in relation to the pandemic provide some evidence that technical indicators are genuinely predictive.

This work presents new tools for evaluating the effectiveness of  policies and practices employed in reducing the impact of the current and future pandemics. This was demonstrated in Sections \ref{Sec4B}  \& \ref{Sec4C} where it was seen that through observations of the weekly MACD indicator one could identify both the peaks of each wave of the pandemic, as well as the onset of subsequent waves of infections. Importantly, reliable short term forecasting can provide potentially lifesaving insights into logistical planning, in particular when and where to allocate additional resources such as hospital staff and equipment.

 \vspace{3mm}

{\bf Acknowledgements.}~ 
This work was completed as part of the MIT PRIMES program. We are grateful to Laura Schaposnik for her thoughtful insights and help, and to Kent Vashaw for their comments on a draft.

\appendix
 \vspace{-3mm} 

\section{Local $p$-values for COVID-19 Analysis}
 \vspace{-3mm}

\begin{figure*}[b!]
\vspace{2cm}
\centerline{
        \includegraphics[width=1.1\linewidth]{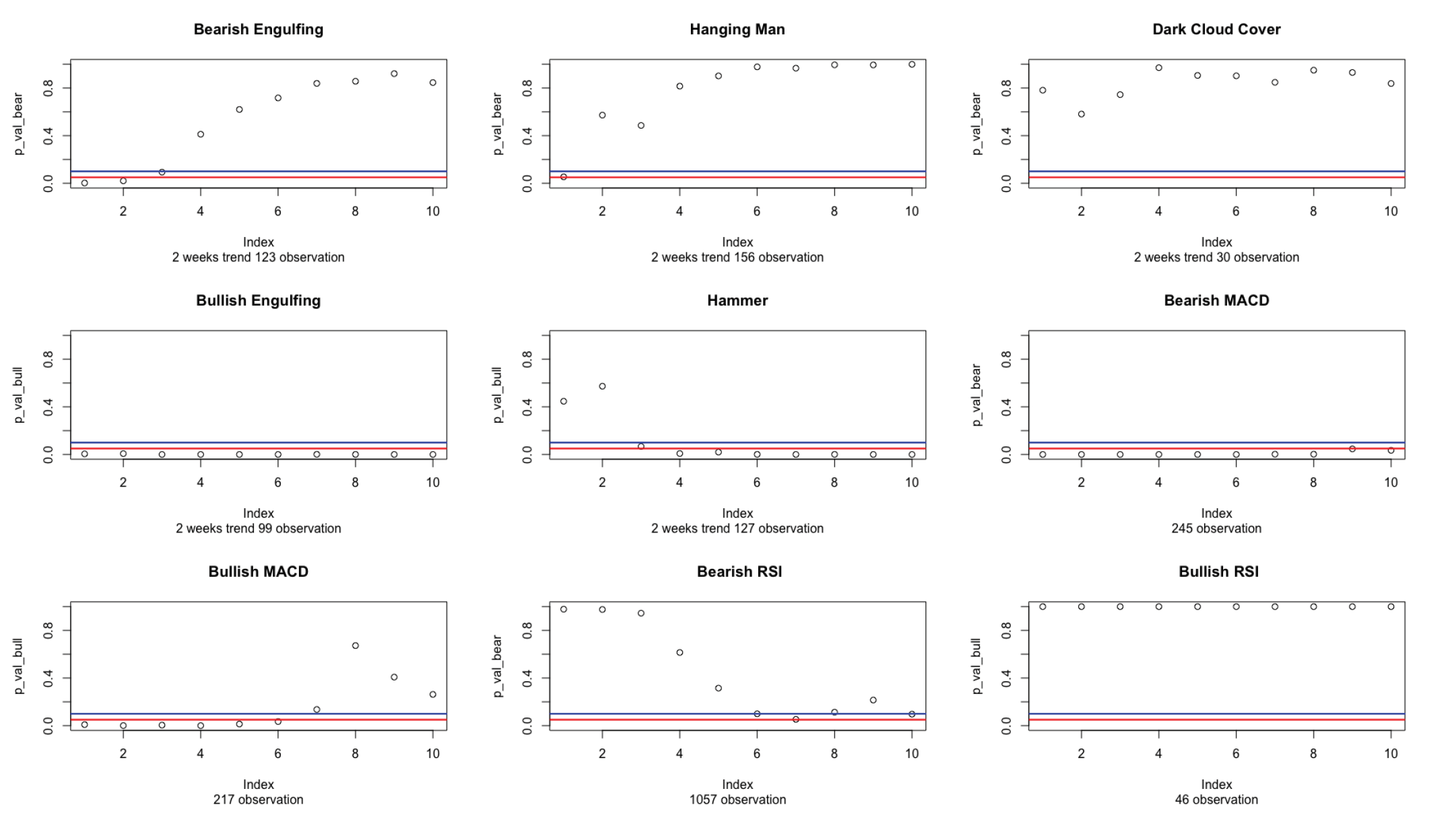} }
    \caption{Plots of (local) $p$-values for specific signals in the pandemic calculated 1 to 10 time periods after a given signal, the horizontal lines indicate the threshold for  $p$-value of 0.05 and 0.10. Multiple points below on of the threshold lines is suggestive of an indicator having predictive power.}
    \label{Fig:result_covid}
\end{figure*}

In this appendix we show how the $p$-value evolves if one changes the point at which the impact of a signal is measured, i.e.~$\Delta t$. However, this implies calculating 90 non-independent $p$-values and thus one should worry about false positives. Such false positives should occur as isolated cases of statistical significance for a particular $\Delta t$ in the case that the indicator is generally not statistically significant. Figure \ref{Fig:result_covid} presents the $p$-values as a function of $\Delta t$, if for most $\Delta t$ the data points fall below either the blue line or the red line, then this implies that the indicators have some predictive power. In the main body we present a more careful study in which we average over independent $p$-values for different $\Delta t$, but this local $p$-value analysis is still insightful, thus we include it here.

 \end{document}